\pdfoutput=1
\documentclass[10pt,twocolumn,letterpaper]{article}

\usepackage[pagenumbers]{cvpr} 

\usepackage{graphicx}
\usepackage{amsmath}
\usepackage{amssymb}
\usepackage{booktabs}
\usepackage{makecell}
\usepackage{subcaption}
\usepackage[export]{adjustbox}
\usepackage[normalem]{ulem}
\usepackage{textpos}  %
\usepackage[misc]{ifsym}
\usepackage{animate}

\usepackage{tabulary,multirow,overpic,xcolor,subfloat}
\usepackage[pagebackref=false, breaklinks=true, letterpaper=true, colorlinks,
citecolor=citecolor, linkcolor=linkcolor, bookmarks=false]{hyperref}
\definecolor{citecolor}{HTML}{0071BC}
\definecolor{linkcolor}{HTML}{ED1C24}

\newcommand{\app}{\raise.17ex\hbox{$\scriptstyle\sim$}}

\newcolumntype{x}[1]{>{\centering\arraybackslash}p{#1pt}}
\newcolumntype{y}[1]{>{\raggedright\arraybackslash}p{#1pt}}

\newlength\savewidth

\makeatletter\renewcommand\paragraph{\@startsection{paragraph}{4}{\z@}
  {.5em \@plus1ex \@minus.2ex}{-.5em}{\normalfont\normalsize\bfseries}}\makeatother

\newcommand\blfootnote[1]{\begingroup\renewcommand\thefootnote{}\footnote{#1}\addtocounter{footnote}{-1}\endgroup}

\DeclareMathAlphabet\mathbfcal{OMS}{cmsy}{b}{n}

\definecolor{Gray}{gray}{0.5}

\newcommand{\modelname}{MagicAvatar\xspace}

\usepackage[capitalize]{cleveref}
\crefname{section}{Sec.}{Secs.}
\Crefname{section}{Section}{Sections}
\Crefname{table}{Table}{Tables}
\crefname{table}{Tab.}{Tabs.}

\def\cvprPaperID{*****}
\def\confName{CVPR}
\def\confYear{2022}

\crefname{section}{\S}{\S\S}
\crefname{subsection}{\S}{\S\S}
\crefformat{table}{Table~#2#1#3}
\crefformat{figure}{Figure~#2#1#3}
\crefformat{equation}{Equation~(#2#1#3)}
\crefformat{algorithm}{Algorithm~#2#1#3}
\crefformat{appendix}{Appendix~#2#1#3}

\newcommand{\authorskip}{\hspace{2.5mm}}

\usepackage{amsmath,amsfonts, mathtools, bm, bbm, nicefrac}

\def\vc{{\bm{c}}}

\def\vm{{\bm{m}}}

\def\vx{{\bm{x}}}
\def\vy{{\bm{y}}}

\DeclareMathAlphabet{\mathsfit}{\encodingdefault}{\sfdefault}{m}{sl}
\SetMathAlphabet{\mathsfit}{bold}{\encodingdefault}{\sfdefault}{bx}{n}

\begin{document}

\title{\modelname:
Multimodal Avatar Generation and Animation}

\author{
 Jianfeng Zhang$^*$ \authorskip 
 Hanshu Yan$^*$ \authorskip
 Zhongcong Xu$^*$ \authorskip 
 Jiashi Feng \authorskip 
 Jun Hao Liew$^*$$^{{\dagger}}$ \\
 ByteDance Inc.\\
 {\small \url{https://magic-avatar.github.io/}}
}

\twocolumn[{
\renewcommand\twocolumn[1][]{#1}
\maketitle
\begin{center}
    \centering
    \vspace{-4mm}
    \animategraphics[width=\textwidth,loop]{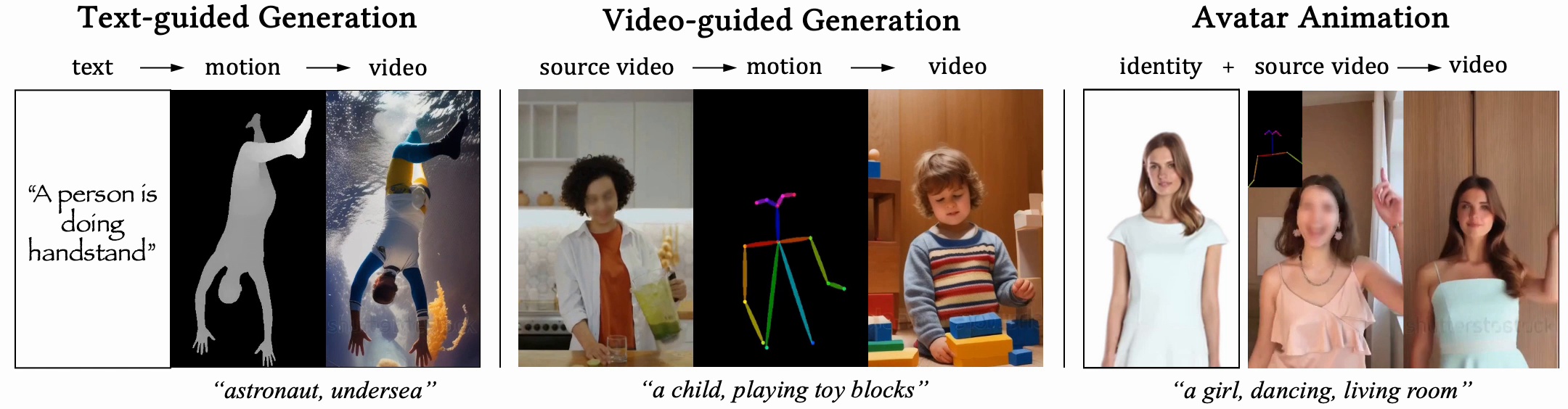}{figs/teaser_frames_compressed/}{1}{16}
    \captionof{figure}
    {We introduce \textbf{\modelname}, a two-stage framework for multimodal avatar generation and animation. 
    \modelname supports a variety of applications, including text-guided avatar generation (1st example), video-guided avatar generation (2nd example) and multimodal avatar animation (3rd example).
    Please note that this figure contains video clips. We encourage readers to \textcolor{magenta}{click and play} using Adobe Acrobat. \textbf{Faces in source videos are blurred} to protect identities.
    }
    \label{fig:teaser}
\end{center}
}]

\blfootnote{$^*$Equal Contribution. $^\dagger$Project Lead.}

\begin{abstract}
This report presents \modelname, a framework for multimodal video generation and animation of human avatars. 
Unlike most existing methods that generate avatar-centric videos directly from multimodal inputs (\eg, text prompts), \modelname explicitly disentangles avatar video generation into two stages: (1) multimodal-to-motion and (2) motion-to-video generation. 
The first stage translates the multimodal inputs into motion/ control signals (\eg, human pose, depth, Densepose); while the second stage generates avatar-centric video guided by these motion signals. 
Additionally, \modelname supports avatar animation by simply providing a few images of the target person. This capability enables the animation of the provided human identity according to the specific motion derived from the first stage.
We demonstrate the flexibility of \modelname through various applications, including text-guided and video-guided avatar generation, as well as multimodal avatar animation. 
\end{abstract}

\vspace{-6mm}
\section{Introduction}

Driven by the rapid advancements in virtual reality, gaming, and social media platforms, the demand for flexible avatar generation and animation has significantly increased over the past years.
However, most existing avatar generation pipelines are time-consuming and complex, posing challenges for novice users. 
This opens the need for more versatile and user-friendly avatar generation tools that can accommodate diverse input modalities, such as text or audio, for content creation.

To address this, we present \modelname, a multimodal framework that transforms multiple input modalities -- text, video, or audio -- into motion/ control signals (\eg, pose, depth, DensePose~\cite{guler2018densepose}, \etc), which are subsequently used to generate or animate an avatar. 
Such two-stage framework is advantageous due to the following reasons:
(1) it eliminates the need for datasets where all modalities co-exist. Instead, \modelname learns to generate motion signals for each modalitiy separately using different paired data (\eg, text-motion, audio-motion, video-motion, \etc), which greatly reduces the difficulty of data collection;
(2) directly learning multimodal-to-video generation is inherently challenging as the model needs to jointly model both the appearance and motion from sparse input signals (\eg, text).
Therefore, explicitly disentangling the avatar generation into two stages helps ease the learning process.

As shown in Fig.~\ref{fig:teaser}, \modelname supports multiple X-to-avatar generation applications, including text-guided and video-guided generation. 
In addition, when users provide images of a specific human subject, \modelname can animate the input subject to follow the motion signals generated in the first stage. 
Despite its simplicity, \modelname's strength lies in its capacity to seamlessly integrate multiple input modalities, enabling customized and realistic avatar generations and animations. By bridging the gap between different input types and the resulting avatar generations, \modelname opens up new possibilities in digital avatar creation and virtual interaction.

\begin{figure}
    \centering
    \includegraphics[width=\linewidth]{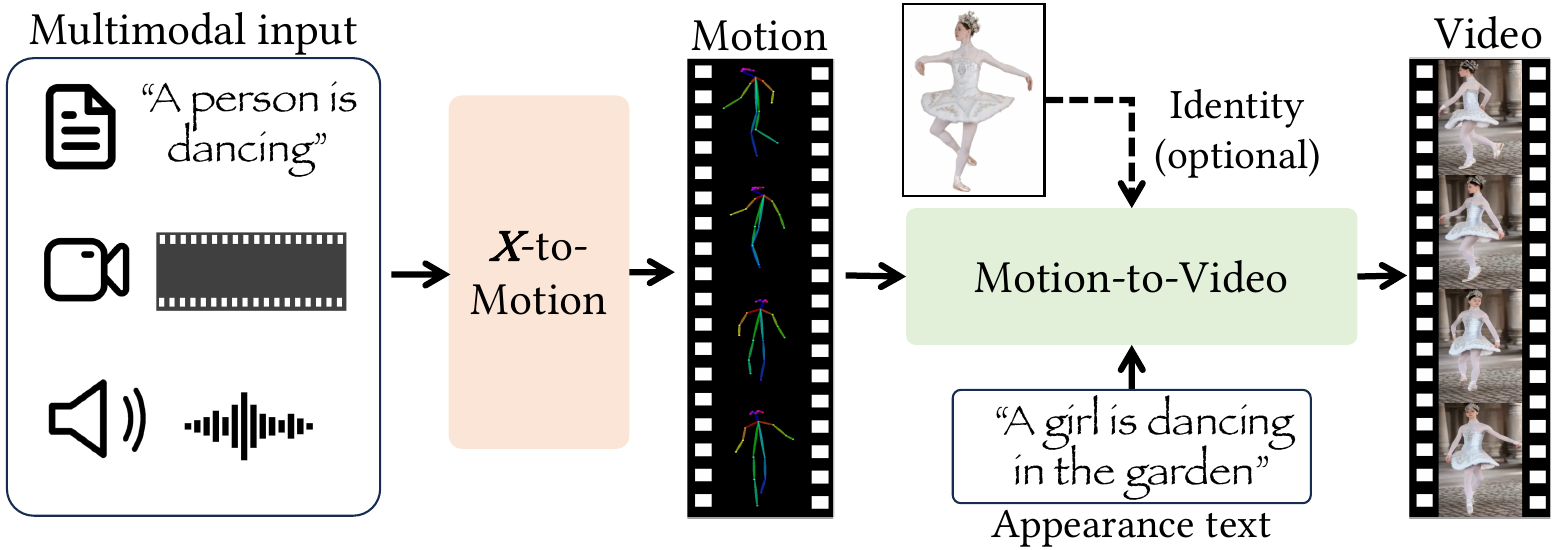}
    \caption{The pipeline of MagicAvatar.}
    \label{fig:enter-label}
\end{figure}

\section{\modelname}

Given a multimodal input $\vy$ (\eg, text, video or audio) specifying a particular motion $\vm$ (\eg, text input ``{\tt a person is doing handstand}") and a prompt description $\vc$ (\eg, ``{\tt astronaut, undersea}"), our goal is to generate a \textbf{realistic} avatar-centric video $\vx=[\vx_1, \cdots, \vx_T]$, where $T$ is the number of frames. The generated video should embody the appearance and background described by the prompt, while adhering to the motion outlined by the multimodal input.
We address this task by explicitly disentangling it into two stages: multimodal-to-motion $f$ and motion-to-video $g$. Mathematically, this can be formulated as:
\begin{equation*} 
    f : \vy \rightarrow \vm, \quad
    g : \vm, \vc \rightarrow \vx.
\end{equation*}
Moreover,  \modelname can inject subject identity into the model for avatar animation. Specifically, 
when users provide images of a specific human subject $I_{\text id}$, 
\modelname encodes the target identity and animates the input human subject according to the motion signals generated in the first stage. 
This process can be formulated as: 
\begin{equation*} 
    g : \vm, \vc, I_{\text id} \rightarrow \vx_{\text id}.
\end{equation*}
In the following, we provide more details about our two-stage pipeline and the identity personalization strategy. The overview of \modelname can be found in Fig.~\ref{fig:enter-label}.

\medskip \noindent \textbf{Multimodal-to-motion generation. }
Our first stage $f$ generates motion signals $\vm$ from multimodal inputs, which can be represented by pose, depth, or DensePose. 
This stage is crucial as it translates the diverse input modalities into a universal `language' that can be used to generate or animate avatar-centric videos.
In this work, we adopt off-the-shelf multimodal-to-motion models for demonstration purpose.
Please refer to Sec.~\ref{sec:application} for more details on the multimodal-to-motion models used.

\medskip \noindent \textbf{Motion-to-video generation. }
The second stage $g$, motion-to-video, aims to produce \textbf{high-fidelity} and \textbf{temporally coherent} videos based on the motion $\vm$ derived from the multimodal inputs $\vy$. 
In this work, we utilize the recently introduced MagicEdit~\cite{magicedit2023} as the core of our motion-to-video generation. Briefly, MagicEdit shows that high-fidelity and temporally consistent video-to-video translation can be achieved by disentangling the learning of content, structure and motion during training. 
Given the generated motion $\vm$ from the first stage, we pass it along with  a prompt description $\vc$ dictating the appearance and background of the video to motion-to-video generation module to produce the final video.

\medskip \noindent \textbf{Identity personalization. }
Beyond generating arbitrary avatars using a prompt $\vc$, 
users may want to to create avatar-centric videos $\vx_{\text id}$ reflecting a specific human identity or animate a given avatar based on the motion sequence from the first stage.
To support this, we can simply train a DreamBooth model~\cite{ruiz2022dreambooth} using a collection of images of the input human subject. 
Specifically, we use 5 images of the target subject $I_{\text id}$ and train for 500 steps
This helps maintain identity consistency throughout the video generation process.

\section{Applications}\label{sec:application}
Next, we explore the potential applications of \modelname, including text-guided avatar generation, video-guided avatar generation, and multimodal avatar animation.

\begin{figure*}
    \centering
    \includegraphics[width=0.97\textwidth]{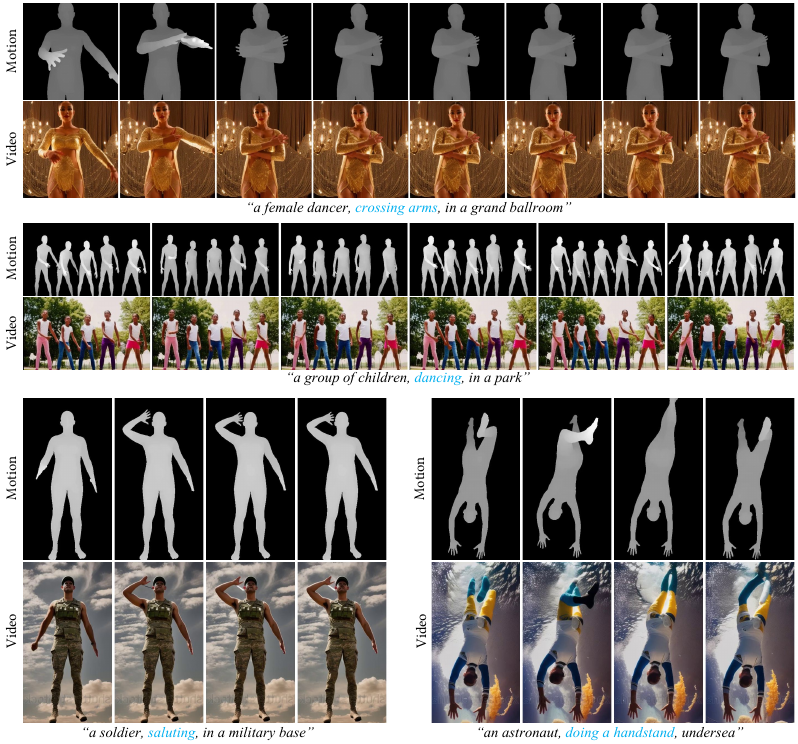}
    \caption{\textbf{Text-guided avatar generation.} \modelname supports avatar generation with simple text prompts only. The prompt expressing the action is highlighted in blue.}
    \label{fig:text-to-video}
\end{figure*}

\medskip \noindent \textbf{Text-guided avatar generation. } 
The two-stage pipeline of \modelname enables the avatar(s) generation using simple text prompts. 
Given a text prompt $\vy_{\text text}$ describing a specific action (\eg, ``{\tt crossing arms}"), we first convert it to motion signals $\vm$ (\eg, human depth or pose) using a pre-trained text-to-motion model MDM~\cite{tevet2023human}. Next, we input the motion signals $\vm$ and a prompt description $\vc$ specifying avatar appearance and video background (\eg, ``{\tt a female dancer, crossing arms, in a grand ballroom}") into the motion-to-video module. This process yields a \textbf{realistic} and \textbf{temporally-coherent} avatar-centric video.
Refer to Fig.~\ref{fig:text-to-video} for examples.

\begin{figure*}
    \centering
    \includegraphics[width=\textwidth]{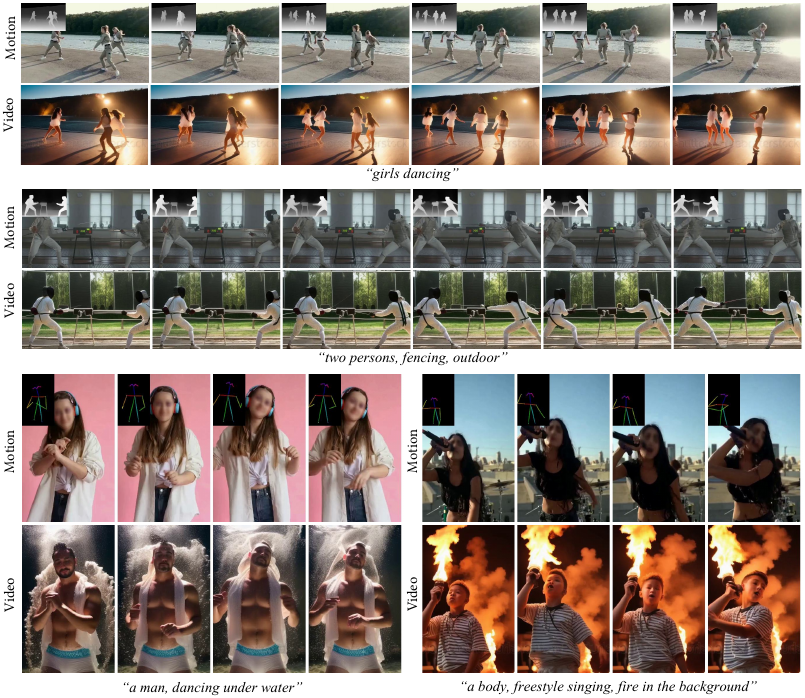}
    \caption{\textbf{Video-guided avatar generation.} We generate new scenes with different subject(s) and different background while following the actions in the source videos. \textbf{Faces in source videos are blurred} to protect identities.
    }
    \label{fig:video-to-video}
\end{figure*}

\medskip \noindent \textbf{Video-guided avatar generation. }
\modelname can also generate avatar videos that mimic the motion in a given source video $\vy_{\text video}$. This application is especially beneficial for creating avatars that perform complex actions or movements that are challenging to describe textually. 
Specifically, we first extract the motion representation (\eg, MiDaS~\cite{ranftl2020towards} for depth maps extraction and OpenPose~\cite{cao2017realtime} human pose estimation) from the source videos. Then, we feed this data into the motion-to-video module to generate an avatar that performs the same actions as in the original video. See Fig.~\ref{fig:video-to-video} for examples.

\begin{figure*}
    \centering
    \vspace{2em}
    \includegraphics[width=\textwidth]{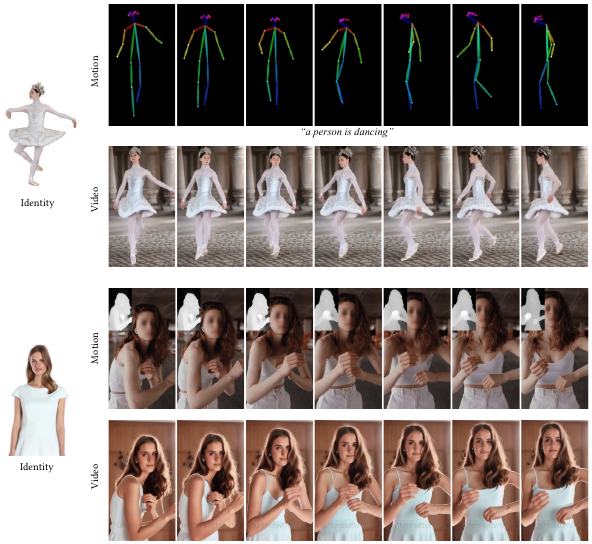}
    \caption{\textbf{Multimodal avatar animation.} Given a few images of the target subject, \modelname allows animating the provided human identity according to the specific motion derived from the first stage (generated from text or extracted from the source video). \textbf{Faces in source videos are blurred} to protect identities.
    }
    \label{fig:animation}
\end{figure*}

\medskip \noindent \textbf{Multimodal avatar animation. }
Our model offers personalization by animating an avatar of a specific subject using various input modalities. Examples can be found in Fig.~\ref{fig:animation}.

\section{Conclusion}
In conclusion,  \modelname demonstrates a significant breakthrough in the realm of avatar generation and animation. By disentangling the process into two distinct stages, \ie, multimodal-to-motion and motion-to-video generation, \modelname provides a flexible and accessible approach to avatar creation. It successfully translates various input modalities into motion signals, which can then be used to generate or animate a realistic, temporally coherent avatar-centric video. Moreover, \modelname's ability to inject a specific human identity into the avatar animation process offers a high level of personalization. Its applications span across text-guided and video-guided avatar generation, as well as multimodal avatar animation, opening up new possibilities in digital avatar creation and virtual interaction. 

{\small
\bibliographystyle{ieee_fullname}
\bibliography{refs}
}

\end{document}